\documentclass{ws-mpla}
\usepackage{slashed,feynmf,epsfig,verbatim,cite}
\allowdisplaybreaks[1]
\renewcommand{\vec}[1]{\boldsymbol{\mathrm{#1}}}

\begin{document}

\title{The running of coupling constants and unitarity in a finite electroweak model}

\author{J. W. Moffat$^{\dag *}$ and V. T. Toth$^{\dag}$\\~\\
$^\dag$Perimeter Institute, 31 Caroline St North, Waterloo, Ontario N2L 2Y5, Canada\\
$^*$Department of Physics, University of Waterloo, Waterloo, Ontario N2L 3G1, Canada}

\maketitle

\begin{abstract}
We investigate the properties of a Finite Electroweak (FEW) Theory first proposed in 1991 \cite{Moffat1991,Clayton1991b,Clayton1991a,Moffat2007f,Moffat2008b}. The theory predicts a running of the electroweak coupling constants and a suppression of tree level amplitudes, ensuring unitarity without a Higgs particle. We demonstrate this explicitly by calculating $W^+_LW^-_L\rightarrow W^+_LW^-_L$ and $e^+e^-\rightarrow W^+_LW^-_L$ in both the electroweak Standard Model and the FEW model.
\end{abstract}

%\pacs{11.10.Lm,12.15.-y,12.60.Cn,13.66.Fg,14.70.Fm}

%11.10.Lm    Nonlinear or nonlocal theories and models
%11.15.Ex    Spontaneous breaking of gauge symmetries
%12.15.-y    Electroweak interactions
%12.60.Cn    Extensions of electroweak gauge sector
%13.66.Fg    Gauge and Higgs boson production in e-e+ interactions
%14.70.Fm    W bosons

\begin{fmffile}{unifigs}

\section{Introduction}

In the Standard Model (SM) of particle physics, the Higgs field, a spin zero field with a nonzero vacuum expectation value is proposed, leading to spontaneous symmetry breaking and a generation of fermion and vector boson masses. Additionally, the Higgs field also resolves the issue of unitarity, by precisely canceling out badly behaved terms in the tree-level amplitude of processes involving longitudinally polarized vector bosons, for instance $W_L^+W_L^-\rightarrow W_L^+W_L^-$ or $e^+e^-\rightarrow W_L^+W_L^-$. The challenge to any theory that aims to compete with the SM without introducing a Higgs particle is to generate the correct fermion and boson masses on the one hand, and ensure unitary behavior for these types of scattering processes on the other.

First proposed in 1991 \cite{Moffat1991}, our Finite Electroweak (FEW) Theory employs three principal ideas \cite{Moffat2008b}:
\begin{itemize}
\item Fermions acquire dynamically generated masses through self-energy \cite{Moffat2007f};
\item A non-local regularization scheme \cite{Evens1991,Cornish1992} ensures that the theory is finite to all orders and does not violate unitarity;
\item A non-trivial choice of the path integral measure factor, consistent with the regularization scheme, leads to spontaneous symmetry breaking and the generation of vector boson masses.
\end{itemize}

The theory is described in detail in a companion paper \cite{Moffat2008b}. In the present work, we focus on a key prediction of the theory: the running of the off-shell $W$ and $Z$ boson masses. This prediction leads to a running of the electroweak coupling constants $g$, $g'$ and $e$, as we demonstrate below. This running of the coupling constants is sufficient to suppress the scattering amplitudes at high energies, ensuring that the amplitudes remain finite and that, as anticipated, unitarity is indeed not violated.

In the first part of this paper, we show how the massive vector boson propagator of the FEW theory leads to a running of the off-shell mass and the consequent running of the electroweak coupling constant. In the second part, we calculate the on-shell rest masses of the $W$ and $Z$ bosons along with the $\rho$ parameter. Lastly, in the third part we demonstrate via explicit calculation how the running of $g$ is sufficient to suppress the scattering amplitude in $W^+_LW^-_L\rightarrow W^+_LW^-_L$ and $e^+e^-\rightarrow W^+_LW^-_L$ processes.

In this paper, we use the metric signature $(+,-,-,-)$, and set $\hbar=c=1$. The Feynmann rules of the FEW theory are listed in the Appendix.

\section{Vector boson masses and the running of the electroweak coupling constant}

In the FEW theory, vector boson propagators take the form \cite{Moffat2008b}:
\begin{equation}
iD_f^{\mu\nu}=-i\left(\frac{\eta^{\mu\nu}-\frac{p^\mu p^\nu}{p^2}}{p^2-\Pi_f^T}+\frac{\frac{\xi p^\mu p^\nu}{p^2}}{p^2-\xi\Pi_f^L}\right),\label{eq:vecprop}
\end{equation}
where $\Pi_f^T$ and $\Pi_f^L$ are the transversal and longitudinal components, respectively, of the vacuum polarization tensor $\Pi_{\mu\nu}$ due to fermion loops, and are given by
\begin{equation}
\Pi_{f\mu\nu}=\Pi_f^T\left(\eta_{\mu\nu}-\frac{p_\mu p_\nu}{p^2}\right)+\Pi_f^L\frac{p_\mu p_\nu}{p^2}.
\end{equation}
It is easy to see that the propagator (\ref{eq:vecprop}) reduces to the SM vector boson propagator if we set $\Pi_f^T=\Pi_f^L=m_V^2$, where $m_V$ is the vector boson rest mass. The gauge-dependent term in the propagator does not contribute to the amplitude and cross section for longitudinally polarized vector bosons, for by definition, the inner product of the 4-momentum and the polarization vector is zero: $\eta^{\mu\nu}p_\mu\epsilon_\nu=0$. In what remains, $\Pi_f^T$ replaces the usual mass squared term in the denominator.

The value of $\Pi_f^T$ has been calculated explicitly in the FEW theory \cite{Moffat2008b} as a function of the four-momentum $q$. For $W^\pm$ bosons:
\begin{equation}
\Pi_{Wf}^T(q^2)=\frac{g_0^2\Lambda_W^2}{(4\pi)^2}\sum_{q^L}(K_{m_1m_2}(q^2)-L_{m_1m_2}(q^2)+2P_{m_1m_2}(q^2)),
\end{equation}
where $q^L$ indicates summation over left-handed fermion doublets with masses $m_1$ and $m_2$, $\Lambda_W$ is the non-local energy scale, and
\begin{align}
K_{m_1m_2}(q^2)=&\int_0^{\frac{1}{2}}d\tau(1-\tau)\bigg[\exp\left(-\tau\frac{q^2}{\Lambda_W^2}-f_{m_1m_2}\right)+\exp\left(-\tau\frac{q^2}{\Lambda_W^2}-f_{m_2m_1}\right)\bigg],\\
P_{m_1m_2}(q^2)=&-\frac{q^2}{\Lambda_W^2}\int_0^{\frac{1}{2}}d\tau\tau(1-\tau)\bigg[E_1\left(\tau\frac{q^2}{\Lambda_W^2}+f_{m_1m_2}\right)+E_1\left(\tau\frac{q^2}{\Lambda_W^2}+f_{m_2m_1}\right)\bigg],\\
L_{m_1m_2}(q^2)=&\int_0^{\frac{1}{2}}d\tau(1-\tau)\bigg[f_{m_1m_2}E_1\left(\tau\frac{q^2}{\Lambda_W^2}+f_{m_1m_2}\right)+f_{m_2m_1}E_1\left(\tau\frac{q^2}{\Lambda_W^2}+f_{m_2m_1}\right)\bigg],
\end{align}
and where
\begin{equation}
f_{m_1m_2}=\frac{m_1^2}{\Lambda_W^2}+\frac{\tau}{1-\tau}\frac{m_2^2}{\Lambda_W^2}.
\end{equation}
The symbol $E_1$ is used for the exponential integral of the first kind, i.e., $E_1(z)=\int_z^\infty e^{-\tau}\tau^{-1}~d\tau$ for $|\mathrm{Arg}(z)|<\pi$.

For $Z$ bosons, the calculation is similar, albeit slightly more complicated:
\begin{align}
\Pi_{Zf}^T(q^2)&=\frac{1}{2}\frac{(g_0^2+g_0'^2)\Lambda_W^2}{(4\pi)^2}\sum_\psi\{[K_{mm}(q^2)-L_{mm}(q^2)]\\
&+P_{mm}(q^2)[2\cos^4\theta_w+32\sin^4\theta_w(Q-T^3)^2-16\sin^2\theta_w\cos^2\theta_wT^3(Q-T^3)]\},\nonumber
\end{align}
where $\psi$ indicates summation over all fermion states (counting left and right-handed fermions and color states separately). As usual, $Q$ and $T^3$ denote the electric charge and weak isospin, respectively, while $\theta_w$ is the Weinberg angle.

\begin{figure}
\centering\includegraphics[width=0.7\linewidth]{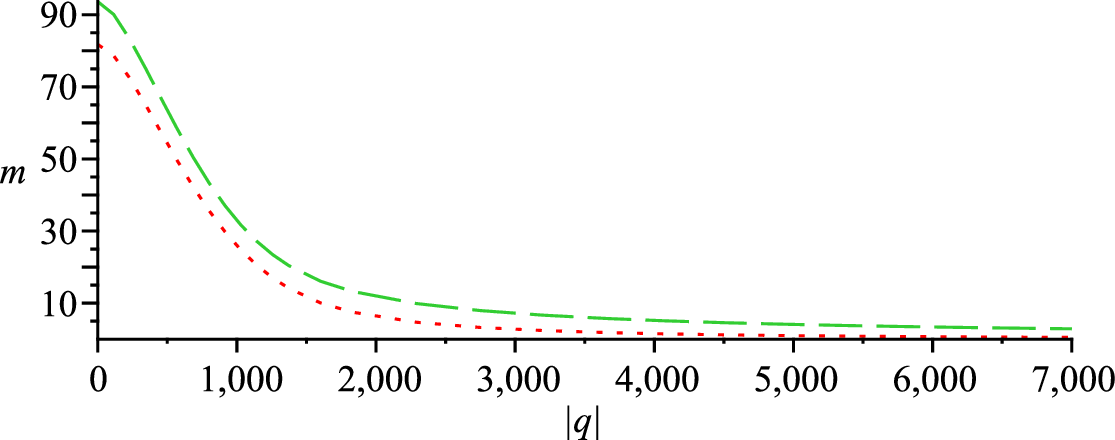}
\caption{The running of the $W$ mass (red dotted line) and $Z$ mass (dashed green line) as functions of momentum. Both axes are measured in units of GeV.}
\label{fig:MWMZ}
\end{figure}
Given the way $\Pi^T$ appears in the vector boson propagator, it is reasonable to make the identification,
\begin{equation}
\Pi_{Wf}^T(q^2)=m_W^2(q^2),~~~~~\Pi_{Zf}^T(q^2)=m_Z^2(q^2).\label{eq:PIWZ}
\end{equation}
The running of the vector boson masses as a function of momentum is illustrated in Fig.~\ref{fig:MWMZ}.

% *** Add new para here
%
%
%

In standard perturbation theory, we compose the free and interaction parts of the theory's Lagrangian as:
\begin{equation}
L=L_0+L_I.
\end{equation}
Instead of diagonalizing $L_0$ and treating the interaction part as a perturbation, we introduce the self-energy Lagrangian $L_{\rm self}$ and split $L$ as\cite{Nambu1961}:
\begin{equation}
L=(L_0+L_{\rm self})+(L_I-L_{\rm self})=L_0'+L'_I.
\end{equation}
We can now define a new vacuum and a complete set of ``quasi-particle'' states for which each particle is an eigenmode of $L_0'$. We now solve $L_{\rm self}$ as a perturbation and determine $L_{\rm self}$ without producing additional self-energy effects. The self-consistent nature of the procedure allows the self-energy to be calculated by perturbation theory with the fields defined by a new vacuum which are already subject to the self-energy interaction.

When we rewrite the theory's Lagrangian according to the above prescription in terms of massive vector bosons \cite{Moffat2007f,Moffat2008b}, the Lagrangian picks up a finite mass contribution from the total sum of polarization graphs:
\begin{align}
\label{massmatrix}
L_m=&\frac{1}{8}v^2g^2[(W^1_\mu)^2+(W^2_\mu)^2]+\frac{1}{8}v^2[g^2(W^3_\mu)^2-2gg'W^3_\mu B^\mu+g^{'2}B^2_\mu]\nonumber\\
=&\frac{1}{4}g^2v^2W^+_\mu W^{-\mu}+\frac{1}{8}v^2(W_{3\mu},B_\mu)\left(\begin{matrix}g^2&-gg'\\-gg'&g^{'2}\end{matrix}\right)\left(\begin{matrix}W^{3\mu}\\B^\mu\end{matrix}\right),
\end{align}
where $W^{\pm}_\mu=(W^1_\mu\mp iW_\mu^2)/\sqrt{2}$  and $v$ is the electroweak symmetry breaking scale (which, in the SM, is the vacuum expectation value of the Higgs scalar). We see that we have the usual symmetry breaking mass matrix in which one of the eigenvalues of the $2\times 2$ matrix in (\ref{massmatrix}) is zero, which leads to:
\begin{equation}
m_W=\frac{1}{2}vg,\quad
m_Z=\frac{1}{2}v(g^2+g^{'2})^{1/2},\quad
m_A=0.\label{eq:mWZ}
\end{equation}

\begin{figure}
\begin{minipage}{0.565\linewidth}\includegraphics[width=1.0\linewidth]{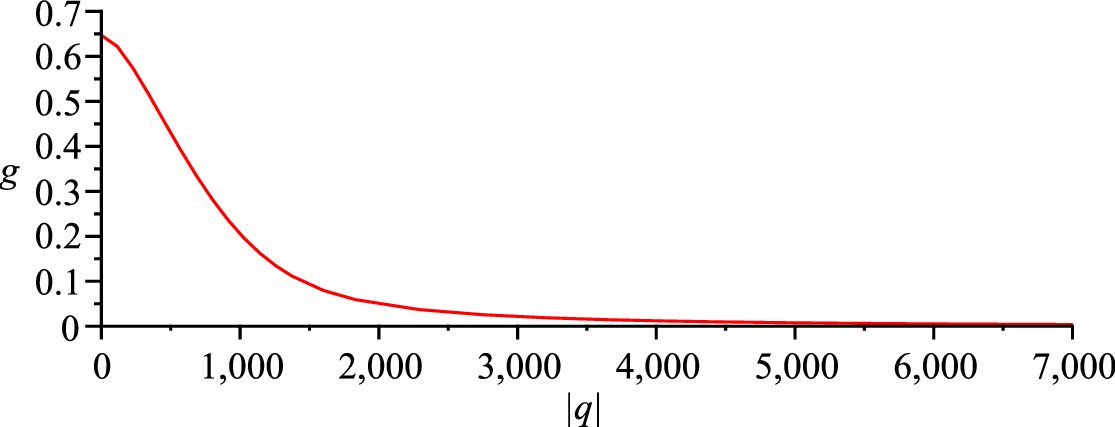}\end{minipage}
\begin{minipage}{0.425\linewidth}\includegraphics[width=1.0\linewidth]{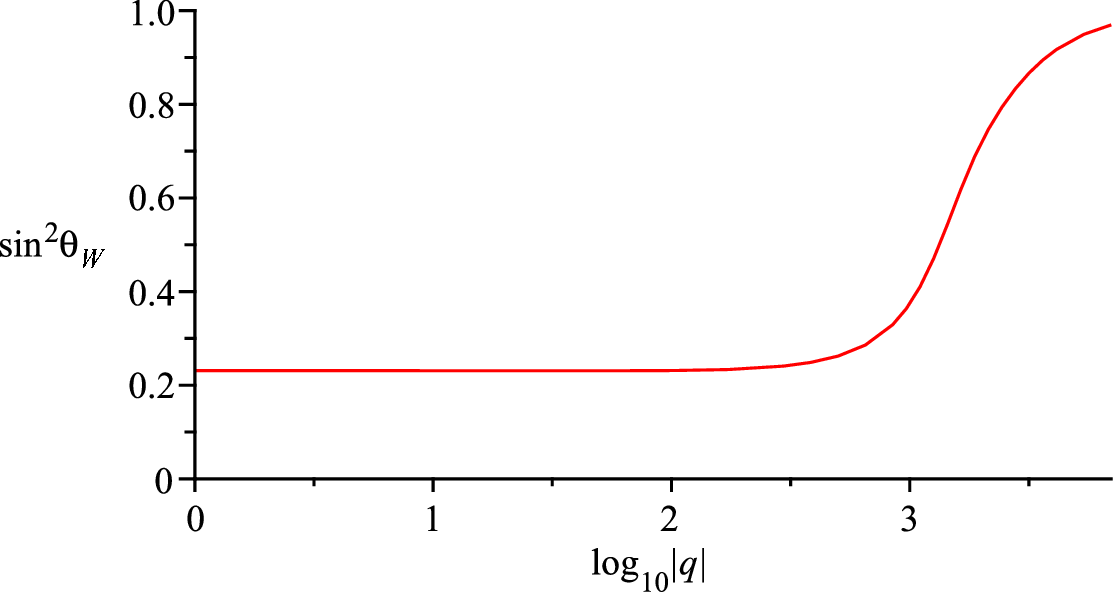}\end{minipage}
\caption{The running of the electroweak coupling constant $g$ (left) as a function of momentum, and the running of the Weinberg angle (right) as a function of momentum. Momentum is measured in GeV.}
\label{fig:gtheta}
\end{figure}

%\begin{figure}
%\centering\includegraphics[width=0.8\linewidth]{g.eps}
%\caption{The running of the electroweak coupling constant $g$ as a function of momentum, measured in GeV.}
%\label{fig:g}
%\end{figure}
%
%\begin{figure}
%\centering\includegraphics[width=0.8\linewidth]{thetaW.eps}
%\caption{The running of the Weinberg angle as a function of momentum, measured in GeV.}
%\label{fig:thetaW}
%\end{figure}

Consistency between (\ref{eq:PIWZ}) and (\ref{eq:mWZ}) can be restored by introducing two running degrees of freedom. Specifically, we choose the constants $g$ and $g'$ to be running in (\ref{eq:mWZ}). Starting with the $W$ mass, we obtain
\begin{equation}
\frac{g^2(q^2)}{g^2(m_Z^2)}=\frac{\Pi_{Wf}^T(q^2)}{\Pi_{Wf}^T(m_Z^2)}.
\label{eq:gg}
\end{equation}
This relationship defines a running of the electroweak coupling constant $g$, which is illustrated in Fig.~\ref{fig:gtheta} (left). Furthermore, from (\ref{eq:mWZ}) and (\ref{eq:gg}), we compute
\begin{equation}
v^2=\frac{4\Pi_{Wf}^T(m_Z^2)}{g^2(m_Z^2)}\simeq (245~\mathrm{GeV})^2.
\end{equation}

In a similar fashion, using the $Z$ mass we obtain
\begin{equation}
\frac{g^2(q^2)+g'^2(q^2)}{g^2(m_Z^2)+g'^2(m_Z^2)}=\frac{\Pi_{Zf}^T(q^2)}{\Pi_{Zf}^T(m_Z^2)},
\end{equation}
which establishes the running of $g'(q^2)$. These relationships also allow us to calculate the running of the Weinberg angle $\theta_w$, which is defined through the ratio of the coupling constants $g$ and $g'$ as
\begin{equation}
\cos\theta_w=\frac{g}{\sqrt{g^2+g'^2}}.
\end{equation}
The running of $\sin^2\theta_w$ is shown in Fig.~\ref{fig:gtheta} (right).

We emphasize that the running of $g$ and $\theta_w$ described in this section arises as a result of the running of the vector boson masses, and does not incorporate radiative corrections, which are the usual origin of the (much more moderate) running of these quantities in the SM.

\section{Vector boson masses and the $\rho$ parameter}

In the previous section, we demonstrated how the rest masses of off-shell vector bosons run with energy. The same formalism can be used to compute the on-shell rest mass of vector bosons by demanding that the following consistency equation be satisfied:
\begin{equation}
m_V^2=\Pi^T_f(m_V^2),~~~~~(V=W,Z).
\end{equation}
In particular, we can postulate this equation for the $Z$-boson ($V=Z$). The left-hand side becomes the $Z$-boson mass, which is known to high accuracy. The expression on the right-hand side involves fermion masses, the electroweak coupling constant, the Weinberg angle, and the non-local energy scale $\Lambda_W$. Of these, all but $\Lambda_W$ are known from experiment; therefore, the equation
\begin{equation}
m_Z^2=\Pi^T_{Zf}(m_Z^2,\Lambda_W),
\end{equation}
where we explicitly indicated the dependence of the right-hand side on $\Lambda_W$, can be solved for $\Lambda_W$. Solving numerically, we obtain
\begin{equation}
\Lambda_W\simeq 541.9~\mathrm{GeV},
\end{equation}
where the precision of $\Lambda_W$ is determined by the accuracy to which $m_Z$ is known, and is not significantly affected by the lack of precision of the known fermion masses, including that of the top quark. With this result at hand, we can move on to the equation for the $W$-boson mass,
\begin{equation}
m_W^2=\Pi^T_{Wf}(m_W^2),
\end{equation}
which we can solve for $m_W$\footnote{The calculation presented in this section is, in effect, the reverse of that performed in \cite{Clayton1991b}; instead of using the known values of $m_W$ and $m_Z$ to obtain $\Lambda_W$ and $m_t$, we use $m_Z$ and $m_t$ to obtain $\Lambda_W$ and $m_W$, and contrast the latter with observation.}, to obtain
\begin{equation}
m_W\simeq 80.05~\mathrm{GeV}.
\end{equation}
The comparable prediction from the SM without radiative corrections is $m_W\simeq 79.95$~GeV. The fact that our predicted value is slightly closer to the observed value of $m_W=80.398\pm 0.025$ \cite{PDG2008} than the SM prediction is consistent with our expectation that radiative corrections, which we have not yet calculated, will be somewhat smaller in magnitude due to suppression by the non-local operator than in the SM.

The relationship between the vector boson masses and the Weinberg angle is customarily captured in the form of the parameter
\begin{equation}
\rho=\frac{m_W^2}{m_Z^2\cos^2\theta_w}.
\end{equation}
Using the observed value of $m_Z=91.1876\pm 0.0021$ \cite{PDG2008}, the computed value of $m_W$, and the Weinberg angle at the $Z$-pole ($\sin^2\theta_w=0.2312$ \cite{PDG2008}), we get the non-trivial result
\begin{equation}
\rho\simeq 1.0023.
\end{equation}
We note, however, that before the computed value of $\rho$ can be compared to observation, radiative corrections must be taken into account, as their magnitude is similar to the difference between our computed value of $\rho$ and the trivial SM tree-level value of $\rho=1$.

\section{$W^+_LW^-_L\rightarrow W^+_LW^-_L$ scattering}

Without the Higgs particle, the SM would violate unitarity in scattering processes that involve longitudinally polarized vector bosons. The scattering of two vector bosons results in a divergent term proportional to $s$. A less rapid divergence, proportional to $\sqrt{s}$, occurs when fermions annihilate into a pair of vector bosons.

In the SM, these divergences are canceled by terms due to tree-level processes that involve the Higgs boson \cite{LlewellynSmith1973,Cornwall1974}. A theory that does not incorporate a Higgs scalar must offer an alternate mechanism to either cancel or suppress the badly behaved terms, in order to maintain unitarity.

The scattering of two longitudinally polarized $W$ vector bosons can take place through one of the following processes:

\begin{equation}
\parbox{1in}{\begin{fmfgraph*}(50,30)
\fmfleftn{i}{2}\fmfrightn{o}{2}
\fmf{boson}{i1,v1}
\fmf{boson}{i2,v1}
\fmf{boson,label=$\gamma/Z^0$,lab.side=left}{v1,v2}
\fmf{boson}{v2,o1}
\fmf{boson}{v2,o2}
\fmflabel{$W^{+}$}{i1}
\fmflabel{$W^{-}$}{i2}
\fmflabel{$W^{+}$}{o1}
\fmflabel{$W^{-}$}{o2}
\end{fmfgraph*}}
+~~~~~~~~~~
\parbox{1in}{\begin{fmfgraph*}(40,40)
\fmfbottomn{i}{2}\fmftopn{o}{2}
\fmf{boson}{i1,v1}
\fmf{boson}{i2,v1}
\fmf{boson,label=$\gamma/Z^0$}{v1,v2}
\fmf{boson}{v2,o1}
\fmf{boson}{v2,o2}
\fmflabel{$W^{+}$}{i1}
\fmflabel{$W^{+}$}{i2}
\fmflabel{$W^{-}$}{o1}
\fmflabel{$W^{-}$}{o2}
\end{fmfgraph*}},\label{eq:WWST}
\end{equation}\vskip 6pt\noindent
in addition to the $4W$ vertex shown in (\ref{eq:4W}).

In order to facilitate an efficient calculation, we use the center-of-mass frame of reference and align the momenta of the incoming particles with the $x_3$-axis. Furthermore, we align the frame such that the scattering takes place in the $x_2x_3$ plane. In this frame of reference, the momenta of the incoming and outgoing particles can be written as
\begin{align}
p_1=&\{E,0,0,|\vec{p}|\},\\
p_2=&\{E,0,0,-|\vec{p}|\},\\
p_3=&\{E,0,|\vec{p}|\sin\theta,|\vec{p}|\cos\theta\},\\
p_4=&\{E,0,-|\vec{p}|\sin\theta,-|\vec{p}|\cos\theta\},
\end{align}
where $E$ is the energy of the incoming $W^-$ particle, $\vec{p}$ is its 3-momentum, and $\theta$ is the scattering angle. In terms of the center-of-mass energy $\sqrt{s}$, these quantities can be expressed as
\begin{align}
E=&\frac{1}{2}\sqrt{s},\label{eq:E}\\
|\vec{p}|=&\sqrt{\frac{s}{4}-m_W^2}.\label{eq:p}
\end{align}

The spatial components of polarization vectors of longitudinally polarized vector bosons are aligned with the respective 3-momenta; the inner product of the polarization vector and the 4-momentum is zero, i.e., $\eta^{\mu\nu}p_\mu\epsilon_\nu=0$. From this, one can compute the unit longitudinal polarization vectors as
\begin{align}
\epsilon_1=&\frac{1}{m_W}\{|\vec{p}|,0,0,E\},\\
\epsilon_2=&\frac{1}{m_W}\{|\vec{p}|,0,0,-E\},\\
\epsilon_3=&\frac{1}{m_W}\{|\vec{p}|,0,E\sin\theta,E\cos\theta\},\\
\epsilon_4=&\frac{1}{m_W}\{|\vec{p}|,0,-E\sin\theta,-E\cos\theta\}.
\end{align}

To facilitate efficient calculation of the vertices in (\ref{eq:WWST}), we define the following function:
\begin{equation}
V_3(\epsilon_1,\epsilon_2,p_1,p_2)=-ig\{(\epsilon_1\cdot\epsilon_2)(p_1-p_2)+\epsilon_2[\epsilon_1\cdot(p_1+2p_2)]-\epsilon_1[\epsilon_2\cdot(2p_1+p_2)]\}.
\end{equation}

Utilizing this function and denoting the photon and $Z$-boson propagators as $D_A^{\mu\nu}$ and $D_Z^{\mu\nu}$, respectively, we can write the matrix element that corresponds to the $s$-channel processes in (\ref{eq:WWST}) as
\begin{align}
i{\cal M}_s=&-iV_3(\epsilon_1,\epsilon_2,-p_1,-p_2)_\mu V_3(\epsilon_4,\epsilon_3,p_4,p_3)_\nu(\sin^2\theta_wD_A^{\mu\nu}+\cos^2\theta_wD_Z^{\mu\nu})\nonumber\\
=&-ig^2\cos\theta\frac{(2m_W^2+s)^2(4m_W^2-s)}{4m_W^4}\left[\frac{\sin^2\theta_w}{(p_1+p_2)^2}+\frac{\cos^2\theta_w}{(p_1+p_2)^2-\Pi_{Zf}^T}\right],
\end{align}
where we used the vector boson propagators of the FEW theory.

The $t$-channel processes in (\ref{eq:WWST}) can be calculated similarly:
\begin{align}
i{\cal M}_t=&-iV_3(\epsilon_1,\epsilon_3,-p_1,p_3)_\mu V_3(\epsilon_4,\epsilon_2,p_4,-p_2)_\nu(\sin^2\theta_wD_A^{\mu\nu}+\cos^2\theta_wD_Z^{\mu\nu})\nonumber\\
=&ig^2 \Bigg[2c_1m_W^2-(2c_1-3)(5c_1-6)\frac{s}{2}+(c_1^2+8c_1-12)(c_1-2)\frac{s^2}{8m_W^2}\nonumber\\
&~~~~~-(c_1-2)^2(c_1+2)\frac{s^3}{32m_W^4}\Bigg]\times\left[\frac{\sin^2\theta_w}{(p_1-p_3)^2}+\frac{\cos^2\theta_w}{(p_1-p_3)^2-\Pi_{Zf}^T}\right],
\end{align}
where we used the shorthand $c_1=\cos\theta+1$.

Finally, we need to evaluate the $4W$ graph (\ref{eq:4W}):
\begin{equation}
i{\cal M}_4=ig^2\left[(c_3^2-12)\frac{s^2}{16m_W^2}-(3c_3-10)\frac{s}{2m_W^2}\right],
\end{equation}
where we used $c_3=\cos\theta+3$.

Summing the matrix elements that we obtained so far, we get
\begin{equation}
i{\cal M}_M=i{\cal M}_s+i{\cal M}_t+i{\cal M}_4=ig^2\left[\frac{(\cos\theta+1)(4m_W^2-3\Pi_{Zf}^T\cos^2\theta_w)}{8m_W^4}s+{\cal O}(1)\right].\label{eq:MM}
\end{equation}

In the SM, $\Pi_{Zf}^T\cos^2\theta_w=m_W^2$ and we get
\begin{equation}
i{\cal M}_M=ig^2\left[\frac{\cos\theta+1}{8m_W^2}s+{\cal O}(1)\right],\label{eq:MM2}
\end{equation}
which clearly violates unitarity for large $s$. However, this behavior is corrected by the addition of the $s$-channel Higgs process:

\begin{equation}
\parbox{1in}{\begin{fmfgraph*}(50,30)
\fmfleftn{i}{2}\fmfrightn{o}{2}
\fmf{boson}{i1,v1}
\fmf{boson}{i2,v1}
\fmf{dashes,label=$H$}{v1,v2}
\fmf{boson}{v2,o1}
\fmf{boson}{v2,o2}
\fmflabel{$W^{+}$}{i1}
\fmflabel{$W^{-}$}{i2}
\fmflabel{$W^{+}$}{o1}
\fmflabel{$W^{-}$}{o2}
\end{fmfgraph*}}.\label{eq:WWH}
\end{equation}
~\par~\par
The associated matrix element becomes
\begin{align}
i{\cal M}_H=&ig^2\bigg\{\sin^2\theta s^3+[m_H^2(\cos^2\theta-2\cos\theta+5)+8m_W^2(\cos\theta-1)]s^2\nonumber\\
&{}+8[m_W^4(3-5\cos\theta)+m_W^2m_H^2(\cos\theta-3)]s+32[m_W^6(\cos\theta-1)+m_W^4m_H^2]\bigg\}\nonumber\\
\bigg/&8m_W^2(s-m_H^2)[(\cos\theta-1)s+4(\cos\theta-1)m_W^2+2m_H^2].
\end{align}
In the high energy limit, we get
\begin{equation}
i{\cal M}_H=-ig^2\left[\frac{\cos\theta+1}{8m_W^2}s+{\cal O}(1)\right],
\end{equation}
canceling out the bad behavior in (\ref{eq:MM2}). The resulting matrix element becomes
\begin{equation}
i{\cal M}_\mathrm{SM}(W^+_LW^-_L\rightarrow W^+_LW^-_L)=ig^2\left[\frac{\cos^2\theta+3}{4\cos\theta_w^2(1-\cos\theta)}-\frac{m_H^2}{2m_W^2}+{\cal O}(s^{-1})\right].\label{eq:WWSM}
\end{equation}

\begin{figure}
\centering\includegraphics[width=0.7\linewidth]{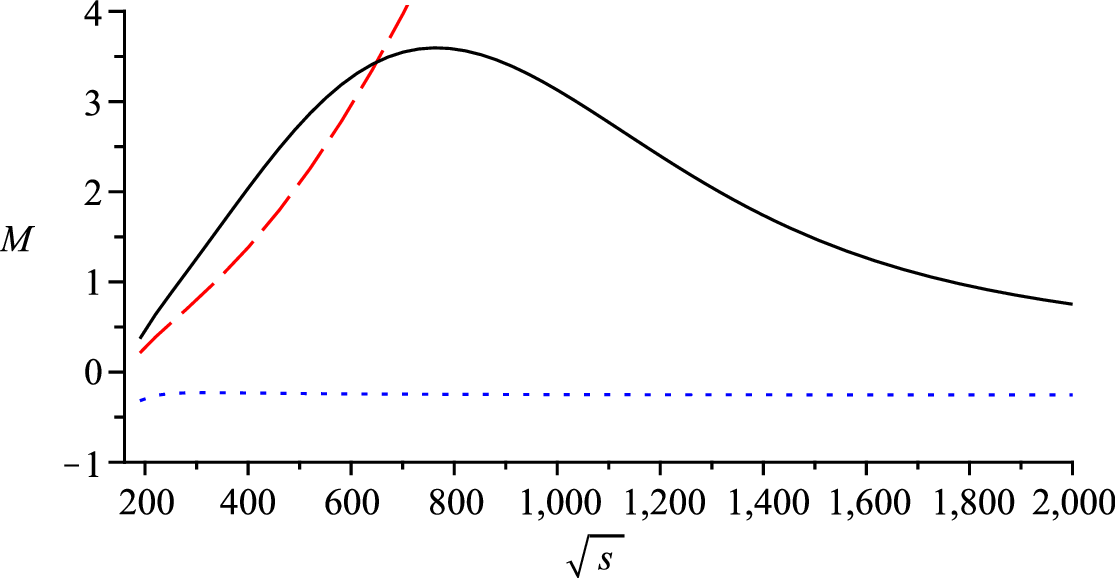}
\caption{The tree-level scattering amplitude of longitudinal $W^\pm$ bosons at a scattering angle $\theta=\pi/2$. The SM (blue dotted line) predicts an asymptotically constant amplitude at high energy. Without the Higgs particle (red dashed line) the amplitude is divergent. In the FEW theory (black solid line) this divergent amplitude is suppressed by the running of the electroweak coupling constant.}
\label{fig:WWWW}
\end{figure}

In the case of the FEW theory, no such additive cancelation takes place. However, the running of the electroweak coupling constant is such that at high $s$, $g(s)s\simeq$~const., which is sufficient to ensure that unitarity is not violated (Fig.~\ref{fig:WWWW}).

Though this calculation is instructive, before the computed amplitudes can be used to make experimentally testable predictions, radiative corrections must be taken into account. Computing the amplitude of $W^+W^-\rightarrow W^+W^-$ scattering with radiative corrections will be a subject of our future research.

\section{$e^+e^-\rightarrow W^+_LW^-_L$ scattering}

The production of $W^+W^-$ pairs from electron-positron collisions can take place via one of the following processes \cite{Peskin1995}:

\begin{equation}
\parbox{1in}{\begin{fmfgraph*}(50,30)
\fmfleftn{i}{2}\fmfrightn{o}{2}
\fmf{fermion}{i1,v1}
\fmf{fermion}{i2,v1}
\fmf{boson,label=$\gamma/Z^0$,lab.side=left}{v1,v2}
\fmf{boson}{v2,o1}
\fmf{boson}{v2,o2}
\fmflabel{$e^{+}$}{i1}
\fmflabel{$e^{-}$}{i2}
\fmflabel{$W^{+}$}{o1}
\fmflabel{$W^{-}$}{o2}
\end{fmfgraph*}}
+~~~~~~~~~~
\parbox{1in}{\begin{fmfgraph*}(40,40)
\fmfbottomn{i}{2}\fmftopn{o}{2}
\fmf{fermion}{i1,v1}
\fmf{boson}{i2,v1}
\fmf{fermion,label=$\nu_e$}{v1,v2}
\fmf{fermion}{o1,v2}
\fmf{boson}{v2,o2}
\fmflabel{$e^{+}$}{i1}
\fmflabel{$W^{+}$}{i2}
\fmflabel{$e^{-}$}{o1}
\fmflabel{$W^{-}$}{o2}
\end{fmfgraph*}}.\label{eq:eeST}
\end{equation}
~\par~\par
To set up the problem, we once again work in the center-of-mass system, with the momenta of the incoming electron and positron coinciding with the $x^3$-axis, and the scattering taking place in the $x^2x^3$ plane. Denoting the 4-momenta of the incoming $e^-$ and $e^+$ with $p_1$ and $p_2$, and those of the outgoing $W^-$ and $W^+$ with $p_3$ and $p_4$, we get
\begin{align}
p_1=&\{E,0,0,|\vec{p}_e|\},\\
p_2=&\{E,0,0,-|\vec{p}_e|\},\\
p_3=&\{E,0,|\vec{p}|\sin\theta,|\vec{p}|\cos\theta\},\\
p_4=&\{E,0,-|\vec{p}|\sin\theta,-|\vec{p}|\cos\theta\},
\end{align}
where $E$ and $\vec{p}$, the center-of-mass energies and $W$ 3-momenta are defined by (\ref{eq:E}) and (\ref{eq:p}), and the electron 3-momentum is
\begin{equation}
|\vec{p}_e|=\sqrt{\frac{s}{4}-m_e^2}.
\end{equation}
Using the Dirac basis for the $\gamma$-matrices, the spinor $u$ corresponding to a given fermion 3-momentum $\vec{p}$ and spin $\sigma$ ($\sigma=\pm1/2$) can be constructed as
\begin{equation}
u(\vec{p},\sigma)=\frac{1}{\sqrt{2}}\begin{pmatrix}A_+-A_-\vec{\sigma}\cdot\vec{p}&\vec{0}\\\vec{0}&A_++A_-\vec{\sigma}\cdot\vec{p}\end{pmatrix}\begin{pmatrix}\chi(\sigma)\\\chi(\sigma)\end{pmatrix},
\end{equation}
where
\begin{equation}
A_\pm=\sqrt{\sqrt{\vec{p}^2+m_e^2}\pm m_e},
\end{equation}
\begin{equation}
\chi(\sigma)=\begin{pmatrix}1/2+\sigma\\1/2-\sigma\end{pmatrix},
\end{equation}
and the matrix-valued vector $\vec{\sigma}$ comprises the three Pauli-matrices. As we are using a metric with $(+,-,-,-)$ signature, the Dirac conjugate of a spinor is given by
\begin{equation}
\bar{u}=u^\dagger\gamma^0.
\end{equation}
Antiparticle spinors are written as
\begin{equation}
v(\vec{p},\sigma)=2\sigma\gamma^5u(\vec{p},-\sigma).
\end{equation}

Denoting the spinors of the $e^-$ and $e^+$ particles with $u$ and $v$, respectively, the $s$-channel annihilation through $\gamma$ and $Z^0$ can be calculated as
\begin{equation}
i{\cal M}_s=-gV_3(\epsilon_3,\epsilon_4,p_3,p_4)_\mu\left[\bar{v}\left(\frac{\gamma^\mu\sin^2\theta_w}{s}+\frac{(1-\gamma^5-4\sin^2\theta_w)\gamma^\mu}{s+\Pi_{Zf}^T}\right)u\right].
\end{equation}

In the limit of large $s$, keeping only terms with powers of $s$ higher than 0, this expression simplifies to
\begin{equation}
i{\cal M}_s=g^2\left[\frac{\sin\theta}{4m_W^2}s+i\frac{m_e\cos\theta}{2m_W^2}\sqrt{s}+{\cal O}(1)\right].
\end{equation}
The term proportional to $s$ in this divergent expression is canceled by the $t$-channel neutrino exchange. This can be written as
\begin{equation}
i{\cal M}_t=-i\frac{g^2}{2}\bar{v}\gamma^\mu\frac{1-\gamma^5}{2}\frac{(\slashed{p}_1-\slashed{p}_3)+m_\nu}{(p_1-p_3)^2-m_\nu^2}\gamma^\nu\frac{1-\gamma^5}{2}u\epsilon^*_{3\mu}\epsilon^*_{4\nu}.
\end{equation}

In the high-energy limit, we get
\begin{equation}
i{\cal M}_t=g^2\left[-\frac{\sin\theta}{4m_W^2}s-i\frac{m_e(1+\cos\theta)}{2m_W^2}\sqrt{s}+{\cal O}(1)\right],
\end{equation}
leading to
\begin{equation}
i({\cal M}_s+{\cal M}_t)=-ig^2\left[\frac{m_e}{2m_W^2}\sqrt{s}+{\cal O}(1)\right].
\end{equation}

In the sum ${\cal M}_s+{\cal M}_t$, the terms proportional to $s$ cancel completely. These terms arise from the interaction of the `proper' helicity components in the incoming electrons: $e^-_Le^+_R$ or $e^-_Re^+_L$. This is consistent with the fact that in the interaction Lagrangian, there are interaction terms combining fermion doublets with fermion doublets, and fermion singlets with fermion singlets, but no interaction terms combine fermion doublets and fermion singlets together.

However, as the electron is massive, its spinor exhibits schizophrenic `wrong' helicity behavior \cite{AH2008}, which leads to the terms proportional to $\sqrt{s}$. These terms do not completely cancel, and unitarity is violated until one introduces the Higgs boson. A fermion-antifermion pair (and, in particular, a lepton-antilepton pair) can annihilate into a Higgs particle, which in turn can decay into a $WW$ pair:

\begin{equation}
\parbox{1in}{\begin{fmfgraph*}(50,30)
\fmfleftn{i}{2}\fmfrightn{o}{2}
\fmf{fermion}{i1,v1}
\fmf{fermion}{i2,v1}
\fmf{dashes,label=$H$}{v1,v2}
\fmf{boson}{v2,o1}
\fmf{boson}{v2,o2}
\fmflabel{$e^{+}$}{i1}
\fmflabel{$e^{-}$}{i2}
\fmflabel{$W^{+}$}{o1}
\fmflabel{$W^{-}$}{o2}
\end{fmfgraph*}}.\label{eq:eeH}
\end{equation}
~\par~\par
The matrix element associated with this process can be written as
\begin{equation}
i{\cal M}_H=ig^2\bar{v}\frac{m_e}{2(s-m_H^2)}u\eta^{\mu\nu}\epsilon^*_{3\mu}\epsilon^*_{4\nu}.
\end{equation}
This matrix element is identically zero if the electron and positron have opposite helicities. However, if they have the same helicity, we obtain
\begin{equation}
i{\cal M}_H=ig^2\frac{m_e}{2m_W^2}\frac{s-2m_W^2}{s-m_H^2}\sqrt{s-4m_e^2}.
\end{equation}
At high $s$, we obtain
\begin{equation}
i{\cal M}_H=ig^2\left[\frac{m_e}{2m_W^2}\sqrt{s}+{\cal O}(1)\right],
\end{equation}
which is exactly the term needed to cancel out the bad $\sqrt{s}$ terms we encountered earlier.

\begin{figure}
\centering\includegraphics[width=0.7\linewidth]{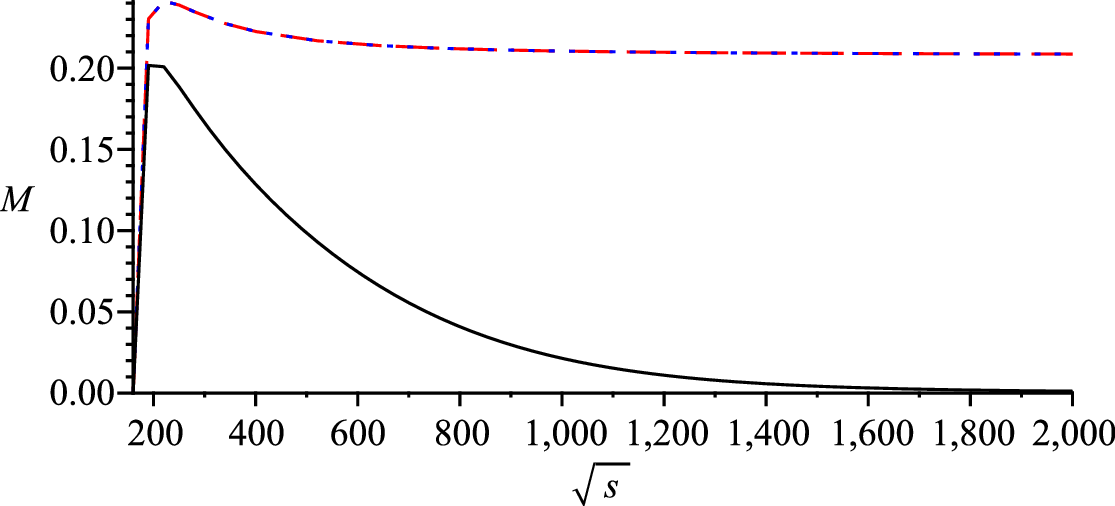}
\caption{The scattering amplitude, at a scattering angle of $\theta=\pi/2$, of electrons and positrons annihilating into longitudinally polarized $W^\pm$ bosons as a function of the center-of-mass energy $\sqrt{s}$, measured in GeV. The SM result (blue dotted line) is indistinguishable from the SM result that was calculated without the Higgs particle (red dashed line), as due to the smallness of $m_e$, the divergent term that is proportional to $m_e\sqrt{s}$ does not begin to dominate until much higher energies. Our Higgless theory, however, predicts a significant suppression of the amplitude even at moderate energies.}
\label{fig:eeWW}
\end{figure}

These results are shown in Fig.~\ref{fig:eeWW}. Although the cancelation due to the Higgs boson is an important feature of the SM, this is not really visible at the energy levels that can be achieved by accelerator, due to the smallness of the electron mass. The FEW theory, however, does predict a significant suppression of the scattering amplitude even at moderate energies, and this may lead to experimentally testable predictions. However, before our predictions can be tested by experiment, it is important to account for radiative corrections, as at specific energies, the magnitude of these corrections may be comparable to the difference between our predictions and those of the SM.

\section{Conclusions}

In this paper, we studied some of the specific consequences of a Higgsless quantum field theory first proposed by Moffat \cite{Moffat1991}. The non-local regularization scheme employed by the theory guarantees unitarity by design; we have seen the explicit manifestation of this effect through the computed running of the electroweak coupling constants and the resulting suppression of divergent amplitudes in processes that involve longitudinally polarized vector bosons. At energy levels that are reachable by either present day accelerators, the LHC, or accelerators that will become operational in the not too distant future, our theory makes predictions that noticeably differ from the predictions of the SM. However, before these predictions can be tested with precision, much work remains. Most importantly, the running of $g$, $\sin^2\theta_w$, and the amplitudes of scattering processes must be calculated with radiative corrections taken into account, in order to remove the inevitable infrared divergences that arise in tree level results at small values of the scattering angle, and to achieve the desired precision in the results.

\renewcommand{\theequation}{A-\arabic{equation}}
\setcounter{equation}{0}

\renewcommand{\thesection}{Appendix~\Alph{section}}
\setcounter{section}{0}

\section{Feynman rules of the Standard Model and the FEW theory}

To investigate $W^+_LW^-_L\rightarrow W^+_LW^-_L$ and $e^+e^-\rightarrow W^+_LW^-_L$ events, we require the following vertex rules and propagators (note that we use the $(+,-,-,-)$ metric signature):

\begin{align}
\parbox{0.65in}{\begin{fmfgraph*}(40,30)
\fmfleftn{i}{1}\fmfrightn{o}{1}
\fmf{dashes,label=$H$,lab.side=left}{i1,o1}
\end{fmfgraph*}}&~~~~\frac{i}{p^2-m_H^2},\\
\parbox{0.65in}{\begin{fmfgraph*}(40,10)
\fmfleftn{i}{1}\fmfrightn{o}{1}
\fmf{fermion,label=$f$,lab.side=left}{i1,o1}
\end{fmfgraph*}}&~~~~i\frac{\slashed p+m_f}{p^2-m_f^2},\\
\parbox{0.65in}{\begin{fmfgraph*}(40,10)
\fmfleftn{i}{1}\fmfrightn{o}{1}
\fmf{boson,label=$\gamma$,lab.side=left}{i1,o1}
\fmflabel{$^\mu$}{i1}
\fmflabel{$^\nu$}{o1}
\end{fmfgraph*}}&~~-\frac{i}{p^2}\left[\eta^{\mu\nu}-(1-\xi)\frac{p^\mu p^\nu}{p^2}\right],\\
\parbox{0.65in}{\begin{fmfgraph*}(40,10)
\fmfleftn{i}{1}\fmfrightn{o}{1}
\fmf{boson,label=$Z^0$,lab.side=left}{i1,o1}
\fmflabel{$^\mu$}{i1}
\fmflabel{$^\nu$}{o1}
\end{fmfgraph*}}&~~-\frac{i}{p^2}\left[\frac{\eta^{\mu\nu}p^2-p^\mu p^\nu}{p^2-\Pi_f^T}+\frac{\xi p^\mu p^\nu}{p^2-\xi\Pi_f^L}\right],\\
&\nonumber\\
\parbox{0.55in}{\begin{fmfgraph*}(40,30)
\fmfleftn{i}{2}\fmfrightn{o}{1}
\fmf{boson}{i1,v1}
\fmf{boson}{i2,v1}
\fmf{dashes,label=$H$,lab.side=left}{v1,o1}
\fmflabel{$W^-_\beta$}{i1}
\fmflabel{$W^+_\alpha$}{i2}
\end{fmfgraph*}}&~~~~igm_W\eta_{\alpha\beta},\\
&\nonumber\\
&\nonumber\\
\parbox{0.45in}{\begin{fmfgraph*}(40,30)
\fmfleftn{i}{2}\fmfrightn{o}{1}
\fmf{boson}{i1,v1}
\fmf{boson}{i2,v1}
\fmf{boson,label=$A_\gamma(q)$,lab.side=left}{v1,o1}
\fmflabel{$W^-_\beta(p_2)$}{i1}
\fmflabel{$W^+_\alpha(p_1)$}{i2}
\end{fmfgraph*}}&\begin{matrix}-ig\sin\theta_w[(p_2-p_1)_\gamma\eta_{\alpha\beta}\\~~~~+(q-p_2)_\alpha\eta_{\beta\gamma}+(p_1-q)_\beta\eta_{\gamma\alpha}],\end{matrix}\\
&\nonumber\\
&\nonumber\\
\parbox{0.45in}{\begin{fmfgraph*}(40,30)
\fmfleftn{i}{2}\fmfrightn{o}{1}
\fmf{boson}{i1,v1}
\fmf{boson}{i2,v1}
\fmf{boson,label=$Z^0_\gamma(q)$,lab.side=left}{v1,o1}
\fmflabel{$W^-_\beta(p_2)$}{i1}
\fmflabel{$W^+_\alpha(p_1)$}{i2}
\end{fmfgraph*}}&\begin{matrix}-ig\cos\theta_w[(p_2-p_1)_\gamma\eta_{\alpha\beta}\\~~~~+(q-p_2)_\alpha\eta_{\beta\gamma}+(p_1-q)_\beta\eta_{\gamma\alpha}],\end{matrix}\\
&\nonumber\\
&\nonumber\\
\parbox{0.55in}{\begin{fmfgraph*}(40,30)
\fmfleftn{i}{2}\fmfrightn{o}{1}
\fmf{fermion}{i1,v1}
\fmf{fermion}{i2,v1}
\fmf{dashes,label=$H$,lab.side=left}{v1,o1}
\fmflabel{$e^-$}{i1}
\fmflabel{$e^+$}{i2}
\end{fmfgraph*}}&~~-ig\frac{m_e}{2m_W},\\
&&\nonumber\\
&&\nonumber\\
\parbox{0.55in}{\begin{fmfgraph*}(40,30)
\fmfleftn{i}{2}\fmfrightn{o}{1}
\fmf{fermion}{i1,v1}
\fmf{fermion}{i2,v1}
\fmf{boson,label=$A_\mu$,lab.side=left}{v1,o1}
\fmflabel{$e^-$}{i1}
\fmflabel{$e^+$}{i2}
\end{fmfgraph*}}&~~~~ig\sin\theta_w\gamma_\mu,\\
&&\nonumber\\
&&\nonumber\\
\parbox{0.55in}{\begin{fmfgraph*}(40,30)
\fmfleftn{i}{2}\fmfrightn{o}{1}
\fmf{fermion}{i1,v1}
\fmf{fermion}{i2,v1}
\fmf{boson,label=$Z^0_\mu$,lab.side=left}{v1,o1}
\fmflabel{$e^-$}{i1}
\fmflabel{$e^+$}{i2}
\end{fmfgraph*}}&~~~~ig\frac{1-\gamma_5-4\sin^2\theta_w}{4\cos\theta_w}\gamma_\mu,\\
&\nonumber\\
&\nonumber\\
\parbox{0.55in}{\begin{fmfgraph*}(40,40)
\fmftopn{o}{1}\fmfbottomn{i}{2}
\fmf{fermion}{i1,v1}
\fmf{boson}{i2,v1}
\fmf{fermion}{v1,o1}
\fmflabel{$e^-$}{i1}
\fmflabel{$W^-_\mu$}{i2}
\fmflabel{$\nu_e$}{o1}
\end{fmfgraph*}}&~~~~-ig\frac{1-\gamma_5}{2\sqrt{2}}\gamma_\mu,\\
&\nonumber\\
&\nonumber\\
\parbox{0.55in}{\begin{fmfgraph*}(40,30)
\fmfleftn{i}{2}\fmfrightn{o}{2}
\fmf{boson}{i1,v1}
\fmf{boson}{i2,v1}
\fmf{boson}{v1,o1}
\fmf{boson}{v1,o2}
\fmflabel{$W^-_\lambda$}{i1}
\fmflabel{$W^+_\mu$}{i2}
\fmflabel{$W^-_\rho$}{o1}
\fmflabel{$W^+_\nu$}{o2}
\end{fmfgraph*}}&~~~~ig^2(2\eta_{\mu\nu}\eta_{\lambda\rho}-\eta_{\mu\lambda}\eta_{\nu\rho}-\eta_{\mu\rho}\eta_{\nu\lambda}).\label{eq:4W}\\
&\nonumber
\end{align}

\end{fmffile}

For the vertices, all momenta are assumed to be pointing inward towards the vertex.

\bibliography{refs}

\end{document}